\documentclass[prl, aps, bibliography, twocolumn, reprint, showpacs, footnoteinbib,superscriptaddress]{revtex4-1}
\usepackage{graphicx}
\usepackage{amssymb}   
\usepackage{amsfonts}
\usepackage{amsmath}
\usepackage{hyperref}
\usepackage{color}
\usepackage{braket}
\usepackage{overpic}
\usepackage{rotating}

\newcommand{\nocontentsline}[3]{}
\newcommand{\tocless}[2]{\bgroup\let\addcontentsline=\nocontentsline#1{#2}\egroup}

\newcommand{\be}{\begin{equation}}
\newcommand{\ee}{\end{equation}}

\begin{document}
	
\title{Quantum many-body dynamics in two dimensions with artificial neural networks}
\author{Markus Schmitt}
\affiliation{Department of Physics, University of California at Berkeley, Berkeley, CA 94720, USA}
\author{Markus Heyl}
\affiliation{Max-Planck-Institut f\"ur Physik komplexer Systeme, N\"othnitzer Stra{\ss}e 38,  01187 Dresden, Germany}
	
\begin{abstract}
The efficient numerical simulation of nonequilibrium real-time evolution in isolated quantum matter constitutes a key challenge for current computational methods.
This holds in particular in the regime of two spatial dimensions, whose experimental exploration is currently pursued with strong efforts in quantum simulators.
In this work we present a versatile and efficient machine learning inspired approach based on a recently introduced artificial neural network encoding of quantum many-body wave functions.
We identify and resolve key challenges for the simulation of time evolution, which previously imposed significant limitations on the accurate description of large systems and long-time dynamics.
As a concrete example, we study the dynamics of the paradigmatic two-dimensional transverse-field Ising model, as recently also realized experimentally in systems of Rydberg atoms.
Calculating the nonequilibrium real-time evolution across a broad range of parameters, we, for instance, observe collapse and revival oscillations of ferromagnetic order and demonstrate that the reached timescales are comparable to or exceed the capabilities of state-of-the-art tensor network methods.
\end{abstract}
	
\maketitle

\paragraph{Introduction.}
In the past two decades the field of nonequilibrium quantum many-body systems has seen a rapid development driven, in particular, by the remarkable progress in experiments \cite{Greiner2002,Kinoshita2006,Georgescu2014,Martinez2016,Choi2016,Gross2017,Jurcevic2017,Bernien2017,Zhang2017,Gaerttner2017,Choi2017,Levine2018,Hild2014,Barredo2018}.
Today, quantum simulators provide access to dynamics in quantum matter with an unprecedented control, which has led to the observation of genuinely nonequilibrium phenomena such as many-body localization~\cite{Schreiber2015oo,Smith2016,Choi2016}, discrete time crystals~\cite{Choi2017,Zhang:2017}, dynamical quantum phase transitions~\cite{Jurcevic2017,Zhang2018o,Flaeschner2018}, or quantum many-body scars~\cite{Bernien2017}.
A particular frontier pushed forward by experiments recently is toward the nonequilibrium dynamics in two-dimensional (2D) quantum many-body systems \cite{Hild2014,2016Labuhn,Barredo2018,2018Bakr}.
The theoretical description of such unitary time evolution, yet, faces severe limitations.
For instance, rapid entanglement growth or the exponential cost of contraction impose strong constraints on tensor network approaches. Nevertheless, considerable progress has been reported to capture transient dynamics~\cite{Murg2007,Zaletel2015,Hauschild2015,Hashizume2018,Czarnik2019,Hubig2019,Hubig20192,Hashizume2019,Kshetrimayum2019}.
Recently, it has been proposed that machine learning techniques might overcome these difficulties by encoding quantum many-body states in artificial neural networks (ANNs)~\cite{Carleo2017}.
Subsequent efforts, however, raised doubts that this approach can enable the investigation of otherwise inaccessible regimes of nonequilibrium quantum dynamics \cite{Czischek2018}.

\begin{figure}[b]
\includegraphics[width=\columnwidth]{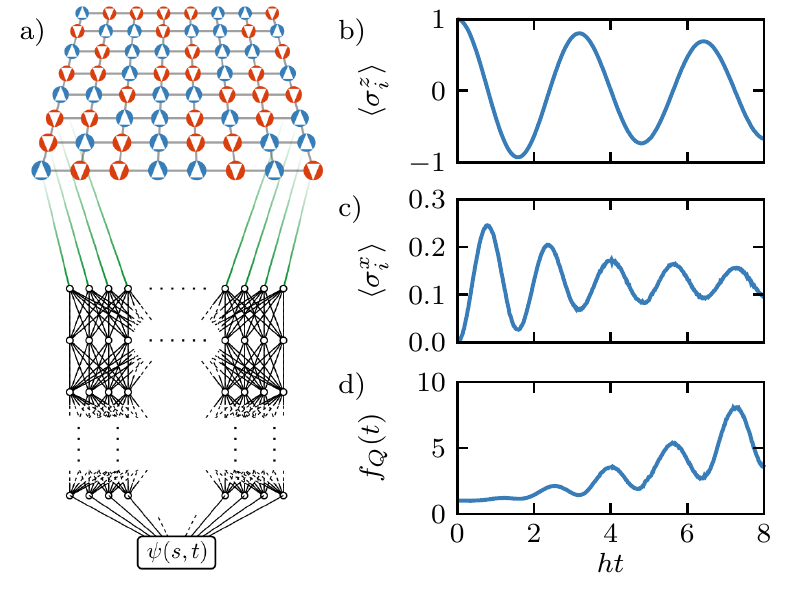}
\caption{(a) Schematic illustration of the artificial neural network (ANN) encoding of many-body wave functions in 2D quantum spin systems. A given spin configuration $s$, blue and red referring to the spin $\uparrow$ and $\downarrow$ state, respectively, functions as the input to an ANN whose output at the end is the corresponding wave function amplitude $\psi_s$.  (b) Collapse and revival of the ferromagnetic order in a quantum Ising model of $8\times 8$ spins on a square lattice after quenching the transverse-field from $h=0$ to $h=2.63h_c$. (c) Dynamics of the transverse magnetization $\langle \sigma_i^x(t) \rangle$. The quantum Fisher information density $f_Q(t)$ in (d) reveals that genuine multipartite entanglement is generated by the unitary evolution.}
\label{fig:coll_rev}
\end{figure}

In this work we overcome hitherto opaque and ultimately prohibitive numerical instabilities of the real-time ANN approach and we thereby expand state-of-the-art capabilities for the simulation of quantum many-body dynamics.
Most importantly, we introduce a novel scheme to obtain a stable solution although only noisy estimates of the variational manifold and it's relation to the physical system are known.
Moreover, we target specific properties of the ANN itself, for instance by utilizing deep architectures, i.e., convolutional neural networks.
They naturally embody the fundamental physical principles of locality and causality, which can enhance the encoding efficiency.
We apply our approach to the paradigmatic transverse-field Ising model on a square lattice, whose nonequilibrium dynamics has recently been shown to be accessible in systems of Rydberg atoms~\cite{2016Labuhn,Barredo2018,2018Bakr}.
With our resulting algorithm we obtain numerically exact results up to timescales comparable to or exceeding the capabilities of current tensor network algorithms, demonstrated by comparison to recent data from infinite Projected Entangled Pair States (iPEPS)~\cite{Czarnik2019}.
Computing the dynamics for a wide range of parameters, we observe, e.g., collapse and revival oscillations of the ferromagnetic order when strongly quenched by a transverse-field, see Fig.~\ref{fig:coll_rev}.
Importantly, we find that at this point the expressivity of the ANN is not the limiting factor and the achieved timescales could be extended at mild polynomial expense.

\paragraph{Neural network wave functions.}
Considering a system of $N$ spin-1/2 degrees of freedom, the quantum many-body wave function can be represented in the basis of spin configurations $s=(s_1,s_2,\dots,s_N)$, $s_j = \uparrow,\downarrow$, as
\begin{equation}
    |\psi\rangle = \sum_s \psi(s)|s\rangle \, .
    \label{eq:amplitudes}
\end{equation}
Because of the exponentially large Hilbert space, wave function based numerical methods aiming at large systems need a strategy to avoid storing the individual amplitudes $\psi(s)$ in memory.
In this work we construct a general-purpose variational wave function $\psi_\eta(s)$, parametrized by $\eta=(\eta_1,\ldots,\eta_M)$, which constitutes an efficient representation of $|\psi\rangle$ if $M$ is much smaller than the Hilbert space size. Being able to provide a good approximation of the amplitudes on the fly ($\psi(s)\approx\psi_\eta(s)$), the variational wave function serves as a generative model, from which we can sample using conventional Monte Carlo techniques. Concretely, the expectation value of any observable $\hat O$ can be obtained as:
\begin{align}
	\braket{\psi_\eta|\hat O|\psi_\eta}=\sum_s|\psi_\eta(s)|^2 O_{\eta}(s)\ ,
\end{align}
with $O_\eta(s)=\sum_{s'}\braket{s|\hat O|s'}\psi_\eta(s')/\psi_\eta(s)$. Since $\braket{s|\hat O|s'}$ is sparse for few-body observables, the expectation value can be computed efficiently by Monte Carlo sampling the probability $p_\eta(s) = |\psi_\eta(s)|^2$; importantly, there is no sign problem associated with this procedure.

Clearly, it might appear difficult to construct a general-purpose generative machine.
However, simple but powerful versions have already been constructed recently for tailored problems~\cite{Schmitt2018,deTomasi2019}.
Aiming for a more versatile approach we now follow the proposal to employ artificial neural networks~\cite{Carleo2017}.
ANNs have the crucial advantage that they are universal function approximators \cite{Hornik1991,Taehwan2003,LeRoux2008}.
As a consequence, any quantum many-body wave function can, in principle, be represented by ANNs provided the network is sufficiently large.
Consequently, the network size acts as a control parameter for our simulations that can be used to check convergence of the results.
Moreover, the celebrated gradient backpropagation algorithm \cite{Dreyfus1962,Rumelhart1985,Mehta2019} enables the efficient numerical treatment of this class of variational wave functions.

As one of the key improvements we propose two modifications of the ANN structure compared to previous works.
First, we explore deep architectures by means of convolutional neural networks, which naturally respect the fundamental principles of locality and causality.
While we provide a detailed description of the CNN wave function in the supplemental material \cite{supplement}, let us point out that CNNs include the Restricted Boltzmann Machines (RBMs), which have been used in previous works for quantum dynamics \cite{Carleo2017,Czischek2018,Fabiani2019,Wu2019}, as the special case of a fully connected single layer CNN with a fixed activation function.
By contrast, CNNs are typically constructed as deep networks with sparse connectivity and arbitrary activation functions.
For ground-state searches, CNN architectures have already been explored previously \cite{Choo2019} with a polynomially enhanced efficiency in encoding entanglement as compared to the RBM \cite{Levine2019}.
The hidden unit density $\alpha$, which specifies the size of an RBM \cite{Carleo2017}, corresponds to the number of channels in terms of a CNN architecture, where the filter diameter $d_F$, that defines the connectivity, equals the linear extent of the system. Accordingly, we will denote the size of a CNN with $L$ layers by a tuple $\alpha=(\alpha_1,\ldots,\alpha_L; d_F)$ with $\alpha_k$ the number of channels in the $k$-th layer \cite{supplement}.

Second, we find that it is crucial for the description of the unitary dynamics to use analytic activation functions for the complex-valued ANNs.
In contrast to ground state searches, which are resilient to the encountering of poles and branch cuts of typical activation functions due to the projective nature of imaginary time evolution, real-time evolution relies on the differentiability of the wave function at any point of the variational manifold in the full complex plane.
In our simulations we use as activation functions a sixth order polynomial in the first layer, which allows us to directly incorporate the $\mathbb Z_2$ symmetry, and odd fifth order polynomials in the following layers to avoid the vanishing gradient problem \cite{Glorot2010,supplement}.

\paragraph{Training and the noisy TDVP.}
Training, i.e., optimizing $\psi_\eta(s)$ to represent the dynamical quantum many-body wave function, is performed by demanding for each time step $\tau$ that the change of parameters $\dot\eta$ minimizes the distance between the time-evolved state $e^{-i\tau H}\ket{\psi_{\eta(t)}}$ and $\ket{\psi_{\eta(t)+\tau\dot\eta}}$ as measured by the Fubini-Study metric $\mathcal D$ \cite{Carleo2017}. The corresponding optimization objective is
\begin{align}
	r^2(t)=\frac{\mathcal D(\ket{\psi_{\eta(t)+\tau\dot\eta}},e^{-i\tau H}\ket{\psi_{\eta(t)}})^2}{\mathcal D(\ket{\psi_{\eta(t)}},e^{-i\tau H}\ket{\psi_{\eta(t)}})^2}\label{eq:tdvp_residual}
\end{align}
where the constant denominator is introduced as a natural scale for $r^2(t)$.
Minimization with respect to $\dot\eta$ yields a first order differential equation for the variational parameters $\eta_k(t)\in\mathbb C$,
\begin{align}
	S_{k,k'}\dot\eta_{k'}=F_{k}\ ,\label{eq:tdvp}
\end{align}
where $S_{k,k'}=\langle\langle O_k^*O_{k'}\rangle\rangle_c$ and $F_k=-i\langle\langle O_k^*E_{\text{loc}}\rangle\rangle_c$ with $k,k'=1,\dots,M$ and $\langle\langle AB \rangle\rangle_c=\langle\langle AB \rangle\rangle-\langle\langle A \rangle\rangle\langle\langle B \rangle\rangle$ a connected correlation function. Here, we introduced the variational derivatives $O_k(s)=\frac{\partial\ln\psi_\eta(s)}{\partial\eta_k}$ and the local energy $E_{\text{loc}}(s)=\sum_{s'}\braket{s|H|s'}\frac{\psi_{\eta}(s')}{\psi_{\eta}(s)}$. The brackets $\langle\langle\cdot\rangle\rangle$ denote expectation values with respect to the normalized probability distribution obtained from $|\psi_\eta(s)|^2$. Notice that Eq.\ \eqref{eq:tdvp} is the well-known TDVP equation, which for holomorphic $\psi_\eta(s)$ equivalently follows from an action principle \cite{Broeckhove1988,Carleo2017,Haegeman2011,Carleo2011}. The Fubini-Study distance \eqref{eq:tdvp_residual} additionally provides us with a practical figure of merit and in the following we will regard the integrated residual, $R^2(t)=\int_0^tdt'r^2(t')$, as a measure of the accuracy of our simulations.
For completeness, we include a derivation of Eq.\ \eqref{eq:tdvp} and the explicit form of the residual \eqref{eq:tdvp_residual} in Ref.~\cite{supplement}.

While solving the exact TDVP equation will yield the optimal parameter update given the variational ansatz, it is important to realize that in practice we will have incomplete knowledge of the equation itself, because both $S_{k,k'}$ and $F_k$ can only be estimated by Monte Carlo sampling. 
In previous works a pseudo-inverse was used to regularize the inversion of the typically ill-conditioned $S$-matrix to avoid contributions from small eigenvalues that can lead to a numerical instability \cite{Carleo2017,Czischek2018,Fabiani2019}. In this work, instead, we follow a different approach by precisely identifying and disregarding the noisy components of Eq.\ \eqref{eq:tdvp}. This new regularization scheme is crucial to be able to reach the network sizes and timescales presented in the following.

For our analysis we consider the TDVP equation \eqref{eq:tdvp} in the eigenbasis of $S$,
\begin{align}
	\sigma_k^2\dot{\tilde\eta}_k=\langle\langle Q_k^*E_{loc}\rangle\rangle_c\equiv\rho_k
\end{align}
where $S_{k,k'}=V_{k,l}\sigma_l^2(V^\dagger)_{l,k'}$, $Q_k=(V^{\dagger})_{k,k'}O_k$, and $\dot{\tilde\eta}_k=(V^{\dagger})_{k,k'}\dot\eta_{k'}$. Our key observation is the fact that the signal-to-noise-ratio of $\sigma_k$, $SNR(\sigma_k)$, is independent of $k$, while $SNR(\rho_k)$ shows a clear $k$-dependence \cite{supplement}. This numerical observation is consistent with the behavior of signal-to-noise-ratios derived analytically by assuming that the joint distribution of $Q_k$ and $E_{loc}$ is Gaussian. In this case, $SNR(\sigma_k)=\sqrt{N_{MC}/2}$ is completely determined by the number of Monte Carlo samples $N_{MC}$, whereas
\begin{align}
	SNR(\rho_k)=\sqrt{\frac{N_{MC}}{1+\frac{\sigma_k^2}{\rho_k^2}\text{Var}(H)}}
	\label{eq:snr}
\end{align}
depends on $k$, and additionally the physical energy variance $\text{Var}(H)$. We find that Eq.\ \eqref{eq:snr} agrees also quantitatively well with the empirically estimated SNR \cite{supplement}.

For our regularization scheme we compute $SNR(\rho_k)$ and discard all components of $\rho_k$, which fall below a fixed threshold. Thereby, we ignore contributions to the TDVP equation of which we have insufficient knowledge due to finite $N_{MC}$. 
Remarkably, the SNR in the Gaussian approximation is directly related to the resulting TDVP residual \eqref{eq:tdvp_residual}: Disregarding the component $k$ increases the residual by $\Delta_k=\frac{|\rho_k|^2}{\sigma_k^2\text{Var}(\hat H)}=SNR(\rho_k)^2/N_{MC}$. 
Hence, increasing $N_{MC}$ at a fixed cutoff will systematically reduce the bias introduced by the regularization; the most important contributions will, however, be accounted for already with small $N_{MC}$ as this result indicates that they have the largest SNR.
Further details of this approach and supporting numerical data are included in the supplemental material \cite{supplement}.

\begin{figure*}[t!]
\includegraphics[width=\textwidth]{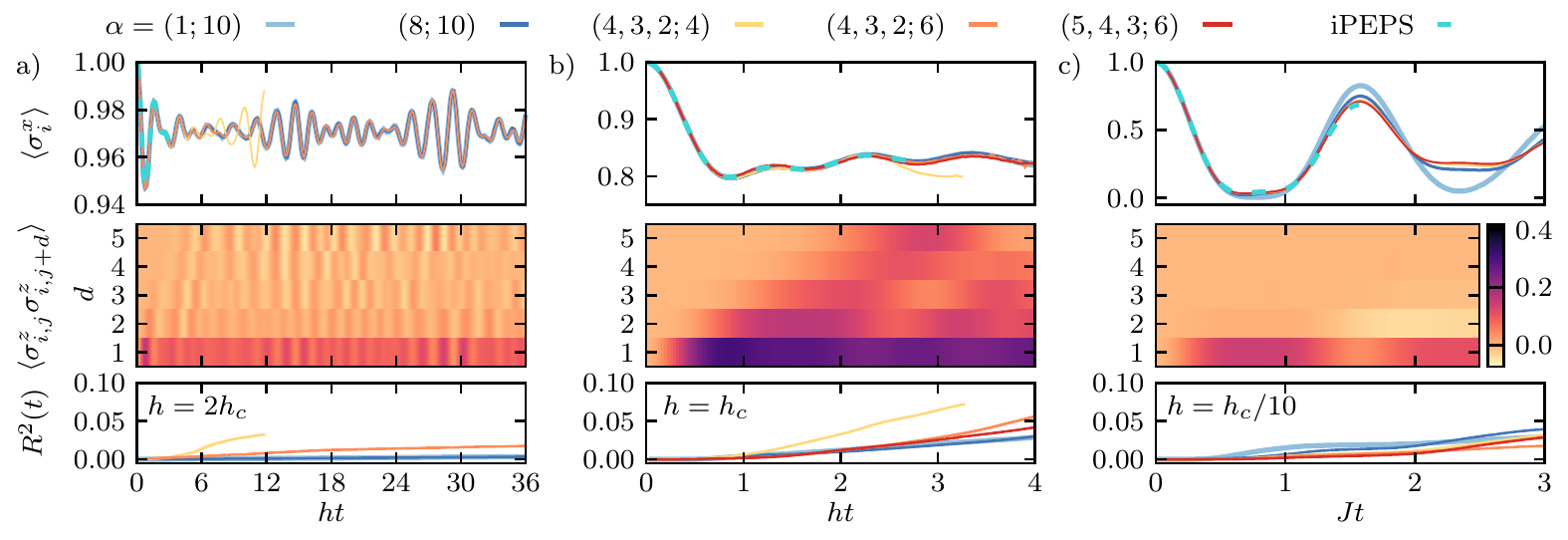}
\caption{Time evolution after quenching a transverse-field Ising model of size $N=10\times10$ from the paramagnetically polarized initial state $\ket{\psi_0}=\ket{\rightarrow}$ a) into the paramagnetic phase at $h=2h_c$, b) to the critical point, and c) into the ferromagnetic phase at $h=h_c/10$. For direct comparison the top row includes data obtained with iPEPS from Ref.\ \cite{Czarnik2019}. The agreement is very good in all cases for the networks with the smallest error $R^2(t)$ (bottom row). The second row shows space-time plots of correlation functions $\braket{\sigma_{i,j}^z\sigma_{i,j+d}^z}$ along the lattice axis from the simulations with minimal error.}
\label{fig:benchmark}
\end{figure*}

\paragraph{Further technical aspects.} 
To propagate the ANN wave function in time we use a second order consistent adaptive integrator. In our implementation we exploit the fact that $S_{k,k'}$ is the metric tensor of the variational manifold \cite{Park2019}, which induces a meaningful measure for the quantification of the integration accuracy \cite{supplement}.

In some cases, when using small network sizes, we found that the details of the resulting dynamics can depend on the initialization of the network. Similar observations have recently been reported in a more general context of training neural networks and it has been proposed to address this issue by ensemble averaging over a number of independently initialized and trained networks \cite{Geiger2019}. We adapted this idea and found that the ensemble average shows good agreement with results from larger networks and the iPEPS reference data \cite{supplement}.

The Monte Carlo sampling from the CNN wave function amplitudes can become computationally very intense, calling for an efficient parallel implementation. Sampling is straightforwardly parallelizable over many processors of a distributed memory machine using a message passing scheme. The network evaluation allows for a shared memory parallelization using the individual cores of a processor or GPUs. Thereby, this machine learning approach allows us to make full use of the computational resources of cutting edge supercomputers for the simulation of quantum many-body dynamics. Further details are contained in the supplemental material \cite{supplement}.

\paragraph{Transverse-field Ising model.}
As a paradigmatic example of a quantum many-body system we consider the transverse-field Ising model on a 2D square lattice, defined by the Hamiltonian
\begin{align}
	H=-J\sum_{\braket{i,j}}\sigma_i^z\sigma_j^z-h\sum_j\sigma_j^x\ .
\end{align}
Here, the $\sigma_i^{x/z}$ denote the Pauli $x$ and $z$ matrices and $\braket{i,j}$ is the set of all neighboring sites in the lattice. The model exhibits a quantum phase transition at the critical transverse-field $h_c/J=3.04438(2)$ \cite{Blote2002} separating a ferromagnetic phase at $h<h_c$ from a paramagnetic phase. This model has recently developed a particular practical relevance, as it is now naturally realized in Rydberg atom quantum simulators \cite{2016Labuhn,Barredo2018,2018Bakr}. Different aspects of its dynamics in 2D have been addressed previously in Refs.\ \cite{Heyl2015,Blass2016,Schmitt2018,Hashizume2018,Czarnik2019,Hashizume2019}.

In the following we demonstrate that the far from equilibrium dynamics induced by quantum quenches can be efficiently simulated using neural network wave functions, independent of the considered parameter regimes.
We choose typical initial conditions of quantum simulators, namely uncorrelated product states $|\psi_0\rangle$.
After preparation the dynamics generated by the Hamiltonian $H$ yields the formal solution $|\psi(t)\rangle = e^{-iHt} |\psi_0\rangle$.

\paragraph{Collapse and revival oscillations.} 
We start with a quench from a ferromagnetically polarized state $\ket{\psi_0}=\ket{\uparrow}=\prod_{l} \ket{\uparrow}_l$ into the paramagnetic phase at $h=2.63h_c$.
The resulting dynamics is shown in Fig.\ \ref{fig:coll_rev} (b-d).
The order parameter $\langle \sigma_l^z\rangle$ exhibits collapse and revival dynamics with decaying amplitude, which is a consequence of relaxation due to interactions.
This is accompanied by the oscillatory buildup of a transverse magnetization.
Notably, significant entanglement is also generated, see Fig.~\ref{fig:coll_rev}(d), where we show the quantum Fisher information density $f_Q(t)=\frac1N\sum_{i,j}\braket{\sigma_i^z\sigma_j^z}_c$.
After two oscillations of the order parameter, $f_Q(t)>8$ implying that genuine multipartite entanglement has been developed of at least $9$ spins~\cite{Hyllus2012,Toth2012}.
We checked the accuracy upon increasing the network size and found that a single layer fully connected CNN with $\alpha=5$ is sufficient for convergence \cite{supplement}.

\paragraph{Quench from a paramagnetic initial condition.}
Next, we consider quenches starting from a paramagnetic initial state $\ket{\psi_0}=\ket{\rightarrow}$.
In this case we can compare our results to data obtained recently with an iPEPS algorithm \cite{Czarnik2019}.

In Fig.~\ref{fig:benchmark} we show results for quenches to weak and strong fields as well as to the quantum critical point, which has previously been identified to constitute a particularly challenging regime for the neural network approach \cite{Carleo2017,Czischek2018}.
For large fields $h_x=2h_c$, we can observe relaxation of the transverse magnetization $\langle \sigma^x(t) \rangle$ to a steady state value with remaining temporal fluctuations due to the finite system size.
In this regime quantum correlations only develop dominantly for nearest-neighboring spins.
For the critical transverse-field $h=h_c$ the magnetization decays to a much smaller value and significant quantum correlations spread in a light-cone fashion also to larger distances indicating a strongly correlated state.
At weak transverse-fields
the dynamics appears more local than in the case of strong fields with quantum correlations emerging almost exclusively between nearest neighbors on the shown timescales. 

Importantly, we find excellent agreement with the dynamics computed using iPEPS for all cases up to the maximally reached times in iPEPS, which are included in Fig.~\ref{fig:benchmark} as dashed lines for comparison.
While iPEPS directly operates in the thermodynamic limit, the utilized machine learning approach enables us to reach significantly larger times for system sizes up to $N=10 \times 10$.
The direct comparison shows that the system size we reach is sufficient to exclude finite-size effects in local observables up to the timescales reached with iPEPS.
To independently assess the accuracy, we perform our simulations with varying network sizes and architectures. 
While fully-connected single-layer CNNs are sufficient to reach convergence on timescales similar to or exceeding iPEPS for quenches into the paramagnetic phase or to the critical point, going to a deep CNN with sparse connectivity yields a substantial improvement over the single-layer network for $h=h_c/10$, indicated also by a significant reduction of the TDVP error $R^2(t)$. In that case, the dynamics remains more local, which can be exploited by using CNNs as we discuss in the supplemental material \cite{supplement}. 
We expect that this feature of deep CNNs can become relevant more generally when addressing larger system sizes, where correlations will remain constrained to smaller fractions of the system extent for longer times.

\paragraph{Discussion.} We have shown that variational time evolution of artificial neural network states constitutes a controlled and accurate approach to simulate dynamics in 2D quantum matter, which is competitive with current state-of-the-art tensor network algorithms. 
An alternative tensor network approach besides iPEPS is based on matrix product states and the approximation of 2D systems using cylindrical geometries \cite{Hashizume2018,Hashizume2019}. For our purpose, however, we chose iPEPS as a reference, because it reflects the full $C_4$ symmetry of the square lattice and we avoid ambiguities caused by boundary effects in the reference data.

The availability of a versatile numerical method for time evolution paves the way to study the nonequilibrium quantum many-body dynamics in 2D and for new benchmarks of quantum simulators against classical simulations. 
The timescales and system sizes presented in this work can be extended at mild polynomial costs; importantly, we demonstrated that the network expressivity is currently not the limiting factor. These findings raise fundamental questions about our understanding of the complexity of quantum states. Moreover, the approach can for example be extended to systems with longer-ranged interactions and without translational invariance, which are challenging to address with tensor network methods.
\tocless{
\begin{acknowledgments}
\phantomsection
We thank P.\ Czarnik, J.\ Dziarmaga, and P.\ Corboz for providing the iPEPS data from their work \cite{Czarnik2019}. Moreover, we acknowledge fruitful discussions with J.\ Budich, M.\ Bukov, H.\ Burau, G.\ Carleo, M.\ Dupont, P.\ Karpov, and M.\ Behr.
MS was supported through the Leopoldina Fellowship Programme of the German National Academy of Sciences Leopoldina (LPDS 2018-07) with additional support from the Simons Foundation. MH acknowledges support by the Deutsche Forschungsgemeinschaft via the Gottfried Wilhelm Leibniz Prize program. Parts of the numerical simulations were performed at the Max Planck Computing and Data Facility in Garching. Moreover, the authors gratefully acknowledge the Gauss Centre for Supercomputing e.V. (www.gauss-centre.eu) for funding this project by providing computing time through the John von Neumann Institute for Computing (NIC) on the GCS Supercomputer JUWELS at Jülich Supercomputing Centre (JSC) \cite{JUWELS}.

\emph{Note added.} Recently, we became aware of related work by I. L\'opez-Guti\'errez and C. Mendl, which appeared simultaneously \cite{Mendl2019}.
\end{acknowledgments}}

\bibliography{literature.bib}

\end{document}


\title{\emph{Supplemental material to} \\Quantum many-body dynamics in two dimensions with artificial neural networks}

\author{Markus Schmitt}
\affiliation{Department of Physics, University of California at Berkeley, Berkeley, CA 94720, USA}
\author{Markus Heyl}
\affiliation{Max-Planck-Institut f\"ur Physik komplexer Systeme, N\"othnitzer Stra{\ss}e 38,  01187 Dresden, Germany}

\maketitle	

\onecolumngrid
\begin{center}
\begin{minipage}{.7\textwidth}
\tableofcontents
\end{minipage}
\end{center}

~
\vspace{1cm}
\twocolumngrid

\section{Convolutional neural network wave function}
\begin{figure}[t]
\includegraphics[width=\columnwidth]{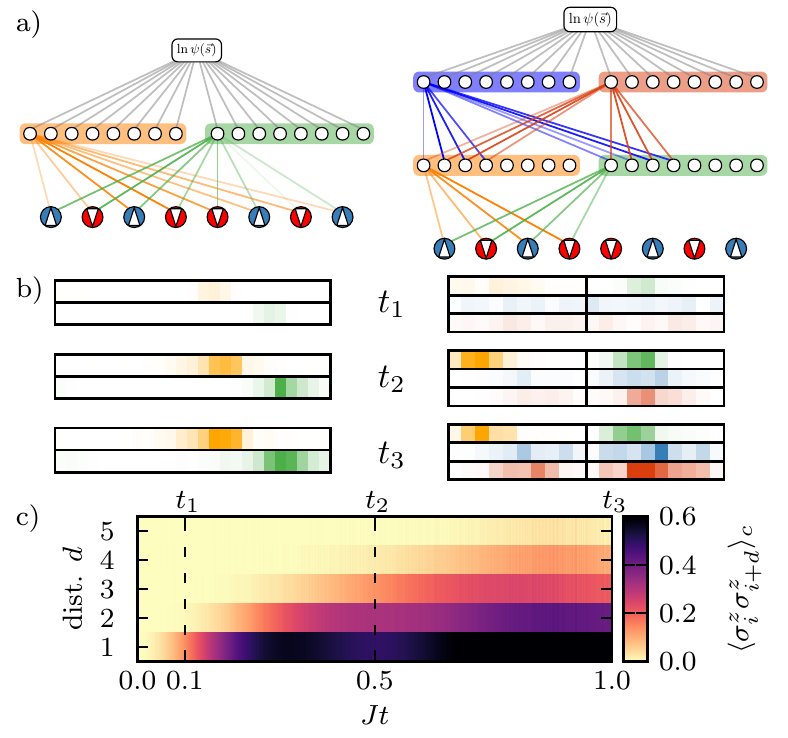}
\caption{Encoding of the hierarchical buildup of correlations in an RBM and a CNN. The network depictions show the distinct couplings, which are repeated in a convolutional manner for all neurons. The intensity plots show the magnitude of the individual couplings color coded matching the network depictions for three different time points $t_1$, $t_2$, and $t_3$. Locality of the physical dynamics is reflected by local build-up of couplings in the RBM, whereas all couplings become involved quickly in the locally connected CNN.}
\label{fig:corr_encoding}
\end{figure}

The convolutional neural network (CNN) is defined as a nested function of alternating non-linear and affine maps, which can be visualized in a layered structure as in Fig.~\ref{fig:corr_encoding}. In this picture, the input layer is filled with the basis configuration $s=(s_1,\ldots,s_N)$ in order to obtain the corresponding coefficient $\psi_\eta(s)$ from the output layer. The term ``convolutional'' owes to the fact that in the CNN architecture the linear part of the affine map actually resembles a convolution of the previous layer with a filter along the orbit generated by the lattice translations. In particular, the values on the vertices, called activations, in the $l$-th layer are obtained as
\begin{align}
	v_{cj}^{(l)}(s)=
	\sigma\Bigg(\sum_{c'=1}^{\alpha_{l-1}}\sum_{k=1}^{N_F}F_{c,c',k}^{(l)}v_{c'\theta_j(k)}^{(l-1)}(s)+b_c^{(l)}\Bigg)\ .
\end{align}
Each layer consists of a number of channels labeled by $c=1\ldots\alpha_l$ as indicated in Fig.~\ref{fig:corr_encoding} by the shaded rectangles in different colors. Each channel has $N_T$ vertices, where $N_T$ is the number of elements $\theta_j$ in the orbit $T$ generated by all lattice translations. The filters of the convolution are denoted by $F_{c,c',k}^{(l)}$ involving $N_F\leq N_T$ activations of each channel $c'$ of the previous layer. The affine map can include a bias $b_c^{(l)}$ for each channel. The filters and biases constitute the set of variational parameters in this ansatz, $\eta\equiv(F,b)$, and they have to be complex numbers in order to encode the complex-valued wave function. Finally, a non-linear activation function $\sigma(\cdot)$ is applied to obtain the values of the activations in the next layer. The recursive relation starts with the initial layer that consists simply of the basis configuration $v_{1j}^{(0)}(s)=s_j=\pm1$.

To obtain a wave function $\psi_\eta(s)$ that is invariant under both translations and point symmetries $\pi\in\mathcal P$ of a lattice, we consider
\begin{align}
	\ln\psi_\eta(s)=\frac{1}{\sqrt{|\mathcal P|\alpha_LN_T}}\sum_{\pi\in\mathcal P}\sum_{c,j}v_{c,j}^{(L)}(\pi(s))\ .
	\label{eq:cnn_def}
\end{align}
Here, we let the CNN encode the logarithm of the wave function coefficients, because in this way the variational derivatives $O_k(s)$ are directly obtained through the backpropagation algorithm.

\begin{table}[t]
\center
\begin{tabular}{|c|c|c|c|}
\hline
Fig.&Sys. size $N$&Network size $\alpha$&Number of parameters $P$\\\hline\hline
1&$8\times8$&$(5;8)$&$320$\\\hline
2&$10\times10$&$(1;10)$&$100$\\
&&$(8;10)$&$800$\\
&&$(4,3,2;4)$&$352$\\
&&$(4,3,2;6)$&$792$\\
&&$(5,4,3;6)$&$1332$\\\hline
\end{tabular}
\caption{Number of parameters $N_P$ of the various networks for the results presented in the main text.}
\label{tab:parameter_numbers}
\end{table}
In all cases presented in this work single layer CNNs are fully connected, whereas for CNNs with multiple layers we utilized filters that connect to square patches of $d_F\times d_F$ neurons in each channel of the previous layer. We refer to $d_F$ as the diameter of the filter. Notice that $d_F\times L$ has to be greater than the linear extent of the system in order to support correlations up to the maximal distances.

As activation function in the first layer we use the first three non-vanishing terms of the series expansion of $\ln\cosh(z)$ around $z=0$, i.e.,
\begin{align}
	\sigma(z)=\frac{z^2}{2}-\frac{z^4}{12}+\frac{z^6}{45}\ .
\end{align}
In subsequent layers we use its derivative
\begin{align}
	\sigma(z)=z-\frac{z^3}{3}+\frac{2}{15}z^5\ .
\end{align}
When testing lower order polynomials we observed that the expressivity of the network is significantly reduced. 
The activation function in the first layer is deliberately chosen as an even polynomial. This allows us to incorporate the $\mathbb Z_2$ symmetry of the model directly in the wave function by setting the biases of the first layer to zero, $b_c^{(1)}\equiv 0$. The non-vanishing derivatives at $z=0$ in the following layers facilitate the mitigation of the vanishing gradient problem, see Ref. [46] of the main text.

\paragraph*{Remark on biases.}  We observe that in the eigenbasis of the $S$-matrix biases mainly contribute to few modes, which are extremely noisy, and therefore cannot be exploited in practice. Therefore, we refrained from including biases in the used networks.

\paragraph*{Network sizes.} The number of variational parameters in a CNN as introduced above is
\begin{align}
	P=N_F\sum_{l=1}^L\alpha_{l-1}\alpha_l+N_B
\end{align}
with $\alpha_0=1$ and $L$ the number of layers in the network. $N_B$ is the total number of biases. In Tab.\ \ref{tab:parameter_numbers} we include the resulting network sizes of the networks used to obtain the data presented in the main text.

\section{Exploiting locality and causality with the CNN}
Figure \ref{fig:corr_encoding} visualizes how the build-up of quantum correlations under time evolution is reflected in the network parameters for an RBM and a CNN. In the interest of clarity we consider the dynamics of a one-dimensional (1D) transverse-field Ising model quenched from the uncorrelated paramagnetic state to the critical point. The hierarchical build-up of correlations is displayed in Fig.\ \ref{fig:corr_encoding}c. The color plots in Fig.\ \ref{fig:corr_encoding}b show snapshots of the magnitudes of the individual couplings of the two networks at three different times, $t_1/J=0.1$, $t_2/J=0.5$, and $t_3/J=1$. In the RBM the build-up of strong correlations at short distances is reflected in localized peaks of the coupling magnitudes of the two channels.
Far from these peaks the couplings vanish, indicating the absence of correlations. 
The vanishing couplings at long distances can be regarded as superfluous, because they could be removed from the ansatz without changing the wave function. Buy contrast, the sparsely connected deep CNN is constructed such that long distance correlations are mediated through the deep layers. 
This means that in the CNN a much larger fraction of the variational parameters plays a significant role despite of the locality of correlations. Since the architecture allows to involve a larger fraction of the available parameters in the encoding of the state, it is potentially more efficient in cases where correlations remain local, such as the quench in Fig. 2c) of the main text.

\onecolumngrid
\section{Time-dependent variational principle}
In the first part of this section we review the derivation of the TDVP equation starting from the Fubini-Study distance. In the second part we discuss in more detail the role of noise in the TDVP equation, including some supporting data.
\subsection{TDVP from Fubini-Study distance}
For the formulation of the time-dependent variational principle we follow Ref. [31] of the main text and consider the distance measured by the Fubini-Study metric between the updated wave function $\ket{\psi_{\eta+\dot\eta\tau}}$ and the one that is obtained by unitary evolution for a short time $\tau$, $e^{-i\hat H\tau}\ket{\psi_\eta}$,
\begin{align}
	\mathcal D\big(\ket{\psi_{\eta+\dot\eta\tau}},e^{-i\hat H\tau}\ket{\psi_\eta}\big)^2
	=
	\arccos\Bigg(\sqrt{\frac{\braket{\psi_{\eta+\dot\eta\tau}|e^{-i\hat H\tau}|\psi_\eta}\braket{\psi_\eta|e^{i\hat H\tau}|\psi_{\eta+\dot\eta\tau}}}{\braket{\psi_{\eta+\dot\eta\tau}|\psi_{\eta+\dot\eta\tau}}\braket{\psi_\eta|\psi_\eta}}}\Bigg)^2
	\equiv\arccos\Bigg(\sqrt{\frac{\braket{\varphi|\phi}\braket{\phi|\varphi}}{\braket{\varphi|\varphi}\braket{\phi|\phi}}}\Bigg)^2
\end{align}
This quantity becomes tractable in the limit of small $\tau$, where we can expand
\begin{align}
	\varphi(s)=\big(1-i\tau E_{\text{loc}}(s)\big)\psi_\eta(s)+\mathcal O(\tau^2)\label{eq:exp1}
\end{align}
and
\begin{align}
	\phi(s)=\big(1+\tau\dot\eta_kO_k(s)\big)\psi_\eta(s)+\mathcal O(\tau^2)\ .\label{eq:exp2}
\end{align}
Notice, that although we will consider a second order consistent expansion of the Fubini-Study metric below, it is sufficient to keep first order terms in the expression above, because the second order terms will cancel each other.
As in the main text, $O_k(s)=\frac{d\log\phi(s)}{d\eta_k}$ denotes the variational derivative, and $E_{\text{loc}}(s)$ is the local energy.

To abbreviate the notation we write
\begin{align}
	\varphi(s)=\big(1-i\tau E_{\text{loc}}(s)\big)\psi_0(s)\equiv\big(1+\mathcal E\big)\psi_\eta(s)
\end{align}
and
\begin{align}
	\phi(s)=\big(1+\tau\dot\eta_kO_k(s)\big)\psi_0(s)\equiv\big(1+R\big)\psi_\eta(s)\ .
\end{align}
In the following $\bar\cdot$ denotes expectation values with respect to $|\psi_\eta(s)|^2$.

Then
\begin{align}
	\frac{\braket{\varphi|\phi}\braket{\phi|\varphi}}{\braket{\varphi|\varphi}\braket{\phi|\phi}}
	&=\frac{\overline{(1+\mathcal E^*)(1+R)}\ \overline{(1+ R^*)(1+\mathcal E)}}{\overline{(1+\mathcal E^*)(1+\mathcal E)}\ \overline{(1+ R)(1+R^*)}}
	\nonumber\\
	&=\frac{1+\oln{\mathcal E^*}+\oln{R}+\oln{\mathcal E^*R}+\oln{R^*}+\oln{\mathcal E}+\oln{R^*\mathcal E}+|\oln{\mathcal E^*}+\oln{R}+\oln{\mathcal E^*R}|^2}
		{1+\oln{\mathcal E}+\oln{\mathcal E^*}+\oln{\mathcal E\mathcal E^*}+\oln{R}+\oln{R^*}+\oln{RR^*}+\big(\oln{\mathcal E}+\oln{\mathcal E^*}+\oln{\mathcal E\mathcal E^*}\big)\big(\oln{R}+\oln{R^*}+\oln{RR^*}\big)}
\end{align}
Now we expand $1/(1+x)=-x+x^2+\mathcal O(x^3)$ and keep only terms of $\mathcal O(\tau^2)$ ($\mathcal E$ and $R$ are of $\mathcal O(\tau)$):
\begin{align}
	\ldots&=\Big(1+\oln{\mathcal E^*}+\oln{R}+\oln{\mathcal E^*R}+\oln{R^*}+\oln{\mathcal E}+\oln{R^*\mathcal E}+\oln{\mathcal E^*}\ \oln{\mathcal E}+\oln{R^*}\ \oln{R}+\oln{\mathcal E^*}\ \oln{R^*}+\oln{\mathcal E}\ \oln{R}\Big)
	\nonumber\\&\quad\quad\times
		\Big(1-\oln{\mathcal E}-\oln{\mathcal E^*}-\oln{\mathcal E\mathcal E^*}-\oln{R}-\oln{R^*}-\oln{RR^*}-\oln{\mathcal E}\ \oln{R}-\oln{\mathcal E}\ \oln{R^*}-\oln{\mathcal E^*}\ \oln{R}-\oln{\mathcal E^*}\ \oln{R^*}
		+(\oln{\mathcal E}+\oln{\mathcal E^*}+\oln{R}+\oln{R^*})^2\Big)+\mathcal O(\Delta^3)
	\nonumber\\
	&=1-\big(\oln{RR^*}-\oln{R}\ \oln{R^*}\big)+\big(\oln{\mathcal ER^*}-\oln{\mathcal E}\ \oln{R^*}\big)+\big(\oln{\mathcal E^*R}-\oln{\mathcal E^*}\ \oln{R}\big)-\big(\oln{\mathcal E^*\mathcal E}-\oln{\mathcal E^*}\ \oln{\mathcal E}\big)+\mathcal O(\tau^3)
\end{align}
Notice that at this point second order terms that we dropped already in Eqs. \eqref{eq:exp1} and \eqref{eq:exp2} would have cancelled the same way as $\oln{\mathcal E}$ and $\oln{R}$ did in the expression above. 

Then the expansion $\arccos(\sqrt{1+x})^2=-x+\mathcal O(x^2)$, yields
\begin{align}
	\mathcal D(\varphi,\phi)^2
	&=\tau^2\Big(\dot\eta_k^*S_{k,k'}\dot\eta_{k'}-F_k^*\dot\eta_k-F_k\dot\eta_k^*+\text{Var}_{\ket{\psi_\eta}}(\hat H)\Big)+\mathcal O(\tau^3)
\end{align}
with
\begin{align}
	S_{k,k'}&=\oln{O_k^*O_{k'}}-\oln{O_k^*}\ \oln{O_{k'}}
	\ ,\quad
	F_k=-i\Big(\oln{O_k^*E_{\text{loc}}}-\oln{O_k^*}\ \oln{E_{\text{loc}}}\Big)
	\ ,\quad
	\text{Var}_{\ket{\psi_\eta}}(\hat H)=\oln{E_{\text{loc}}E_{\text{loc}}^*}-\oln{E_{\text{loc}}}\ \oln{E_{\text{loc}}^*}
\end{align}
Finally, requiring stationarity of $\mathcal D(\varphi,\phi)^2$ with respect to $\dot\eta^*$ up to $\mathcal O(\tau^3)$ yields the linear TDVP equation
\begin{align}
	S_{k,k'}\dot\eta_{k'}=F_k
\end{align}

Again following Ref. [31] of the main text, we estimate the TDVP error as
\begin{align}
	r^2(t)=\frac{\mathcal D\big(\ket{\psi_{\eta+\dot\eta\tau}},e^{-i\hat H\tau}\ket{\psi_\eta}\big)^2}{\mathcal D\big(\ket{\psi_{\eta}},e^{-i\hat H\tau}\ket{\psi_\eta}\big)^2}
	=1+\frac{\dot\eta_k^*R_k-F_k^*\dot\eta_k}{\text{Var}_{\ket{\psi_\eta}}(\hat H)}
\end{align}
with the residual vector $R_k=S_{k,k'}\dot\eta_{k'}-F_k$. This expression is again obtained via a second order consistent expansion in powers of $\tau$.

\paragraph*{Remark on energy conservation.} The TDVP equation defines a Hamiltonian dynamics on the variational manifold, which conserves energy. This fact is readily confirmed by computing the time derivative of the energy expectation value,
\begin{align}
	\frac{d}{dt}\frac{\braket{\psi|\hat H|\psi}}{\braket{\psi|\psi}}=-2\text{Im}\big(\dot\eta_k^*F_k\big)
	=-2\text{Im}\big(F_{k'}^*S^{-1}_{k,k'}F_k\big)=0\ .
\end{align}
Here, the last equality is due to the fact that $S$ is a hermitian matrix. This means, however, that given an accurate estimate of $F_k$ and a sufficiently small time step $\tau$, energy will be conserved in the numerical simulation for \emph{any} hermitian $S^{-1}$. In particular, regularizations of the inversion will not affect energy conservation as long as hermiticity is preserved.

\begin{figure*}[b]
\vspace{-.5cm}
\includegraphics[width=.95\textwidth]{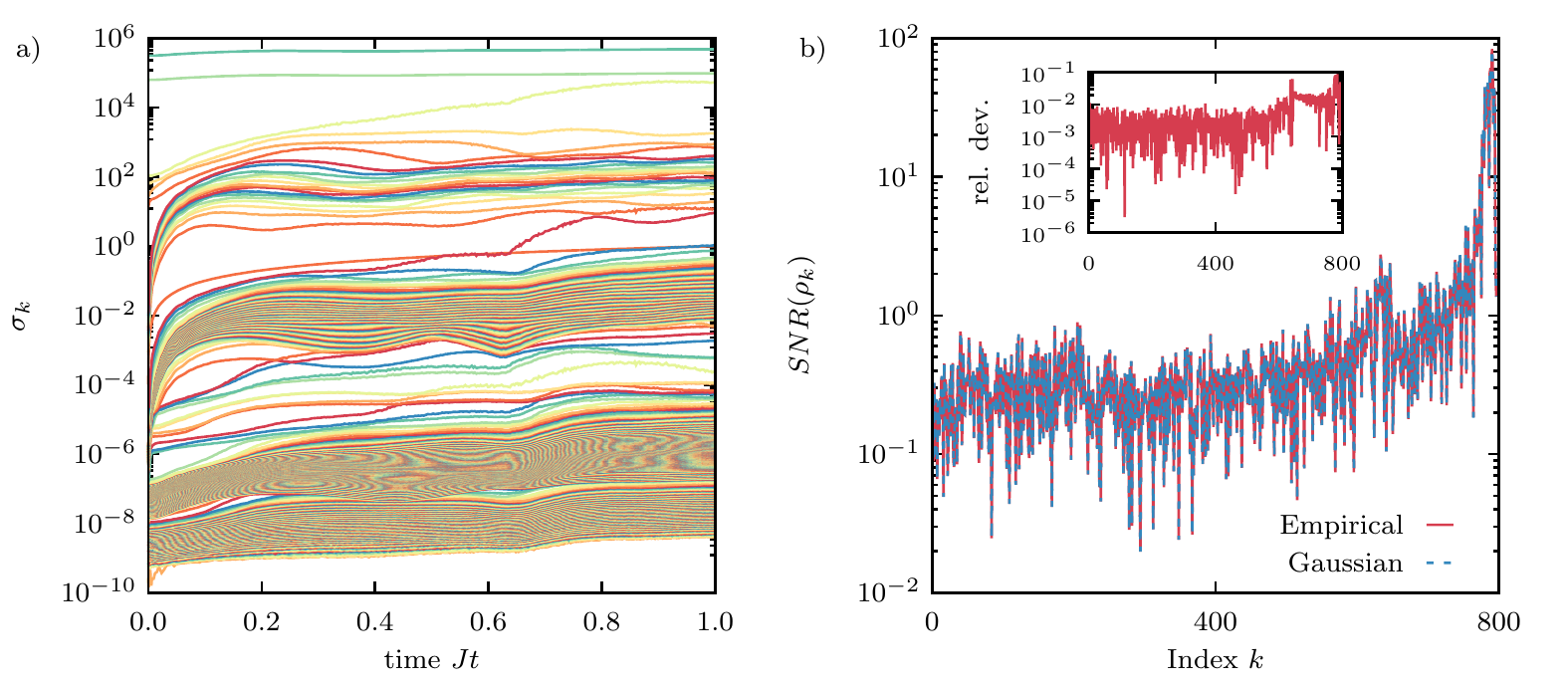}
\caption{The role of noise in the different components of the TDVP equation, vizualized using the example of the quench to the critical point with network $(4,3,2;6)$. (a) Evolution of the spectrum of the $S$-matrix as function of time. Each line corresponds to the evolution of an individual eigenvalue of $S$. The relative amplitude of fluctuations does not depend on the magnitude of $\sigma_k$. Individual eigenvalues and gaps are resolved even very low in the spectrum. (b) Signal-to-noise ratio of $\rho_k$ at $Jt=1$. The $SNR$ shows a clear $k$-dependence and the empirical estimate computed from the observed variance of $Q_k^*\Elocb$ agrees very well with the Gaussian result computed according to Eq.\ \eqref{eq:snr}. The inset shows the relative deviations of both results.}
\label{fig:snr}
\end{figure*}

\newpage
\twocolumngrid
\subsection{The role of noise for the time-dependent variational principle}
For simplicity of notation we use in the following $\bar O_k=O_k-\langle O_k\rangle$ and $\Elocb=\Eloc-\langle\Eloc\rangle$, with which
\begin{align}
	S_{k,k'}=\langle\bar O_k^*\bar O_{k'}\rangle\quad\text{and}\quad F_k=-i\langle\bar O_k^*\Elocb\rangle
\end{align}

An eigenvalue decomposition yields
\begin{align}
	S=V\Sigma^2 V^\dagger
\end{align}
where columns of $V$ hold eigenvectors of $S$ and $\Sigma^2$ is diagonal with $\Sigma_{kk}^2=\sigma_k^2$. We transform the random variables $O_k$ into the eigenbasis and obtain new random variables
\begin{align}
	Q_k=\big(V^\dagger\big)_{k,k'}\bar O_{k'}
\end{align}
with vanishing mean and covariance, while the variance is $\langle |Q_k|^2\rangle=\sigma_k^2$. Thereby, we can also rewrite the TDVP equation
\begin{align}
	\sigma_k^2\big(V^\dagger\big)_{k,k'}\dot\eta_{k'}=-i\langle Q_k^*\Elocb\rangle\equiv\rho_k
\end{align}
As we show in the following, this basis is particularly well suited to analyze the role of Monte Carlo noise in the TDVP equation. Besides numerical data we present analytical forms of the signal-to-noise-ratio of the key quantities, which are based on a Gaussian approximation that will be justified in the last part of this section.

\subsubsection{Noise in the $S$-matrix}
We compute the variances $\sigma_k^2$ from an MC estimate of the matrix $S$. A concern might therefore be, that small $\sigma_k^2$ are more prone to MC fluctuations than large $\sigma_k^2$. This is, however, not the case if the $Q_k$ behave like Gaussian random variables: Assuming the Gaussian property $\langle |Q_k|^4\rangle=3\sigma_k^4$, the signal to noise ratio of $\sigma_k^2$ is 
\begin{align}
	SNR(\sigma_k^2)&=\frac{\sigma_k^2}{\sqrt{\big(\langle |Q_k|^4\rangle-\sigma_k^4\big)/N_{MC}}}
	\nonumber\\
	&=\frac{\sigma_k^2\sqrt{N_{MC}}}{\sqrt{2\sigma_k^4}}=\sqrt{\frac{N_{MC}}{2}}\ ,
\end{align}
i.e., independent of the magnitude of $\sigma_k^2$. This is consistent with observations from our simulations, where the relative MC fluctuations are the same over all orders of magnitude, see, e.g., Fig.\ \ref{fig:snr}.

\subsubsection{Noise in the $F$-vector}
Again assuming a Gaussian for the joint distribution of $Q_k$ and $\Elocb$, the variance of $Q_k^*\Elocb$ is
\begin{align}
	\langle |Q_k^*\Elocb|^2\rangle-|\rho_k|^2=|\rho_k|^2+\sigma_k^2\ \text{Var}(\hat H)
	\label{eq:var}
\end{align}
Therefore, the signal to noise ratio of the right hand side is
\begin{align}
	SNR\big(\langle Q_k^*\Elocb\rangle\big)&=\frac{|\rho_k|}{\sqrt{\big(|\rho_k|^2+\sigma_k^2\ \text{Var}(\hat H)\big)/N_{MC}}}
	\nonumber\\
	&=\frac{\sqrt{N_{MC}}}{\sqrt{1+\frac{\sigma_k^2}{|\rho_k|^2}\text{Var}(\hat H)}}
	\label{eq:snr}
\end{align}
Hence, the signal to noise ratio of the MC estimate on the right hand side of the equation has a dependence on the index $k$; in particular, it depends on the ratio of $\sigma_k^2$ and $|\rho_k|^2$. In Fig.\ \ref{fig:snr} we show a representative example of $SNR(\rho_k)$ as obtained from Eq.\ \eqref{eq:snr} in comparison with the SNR obtained from an explicit Monte Carlo estimate of the variance $\langle |Q_k^*\Elocb|^2\rangle-|\rho_k|^2$. Although $SNR(\rho_k)$ ranges over four orders of magnitude, the relative deviations of the two estimates are never larger than $10\%$. Hence, Eq.\ \eqref{eq:snr} provides a decent quantitative approximation despite the ad-hoc assumption of a Gaussian distribution.

Notice that once we diagonalized the $S$-matrix, the SNR of the r.h.s. only depends on quantities that we know. Hence, we can choose a threshold for the SNR to determine the suited cutoff for the pseudo-inverse based on the data that we have.

\subsubsection{Joint distribution of $Q_k$ and $\Elocb$}
While we cannot present a strict proof we would like to motivate heuristically the origin of the consistency of the observed SNRs with a joint Gaussian distribution of $Q_k$ and $\Elocb$. To this end, we conjecture that the summation of many essentially random terms in both quantities approximately gives rise to a central limit theorem. In particular, the local energy for a given configuration $s$ is a sum of $N$ realizations of random numbers, which are identically distributed due to translational symmetry. Similarly, due to the symmetrization in Eq.\ \eqref{eq:cnn_def} $Q_k(s)$ contains the summation over $N$ random terms.

\section{Details of the numerical procedure}
\subsection{Computational basis}
We always choose the quantization axis depending on the polarization of the initial state: For the $x$-polarized initial state we use the $z$-basis and for the $z$-polarized state we choose the $x$-basis. This facilitates initialization, because networks with small random weights are already close to the desired initial states. Moreover, the $z$-polarized state in the $z$-basis is pathological for computational purposes: For $\psi=\otimes_j\ket{\uparrow}_j$ the $S$-matrix in the TDVP equation vanishes and the dynamics can only be initiated by including higher order terms.
\subsection{Network initialization}
\paragraph*{Single layer.} We initialized single layer networks by drawing random weights from the uniform distribution on $[-w,w]$ with $w=10^{-3}$.
\paragraph*{Deep networks.} For deep networks a suited choice of initial weights is crucial due to the vanishing or exploding gradient problems.
To obtain well-behaved gradients we follow the analysis of Ref. [46] of the main text and draw initial values of $F_{c,c',k}^{(l)}$ from uniform distributions on $[-w^{(l)},w^{(l)}]$ with $w^{(l)}=(N_F(\alpha_{l-1}+\alpha_l))^{-1/2}$.
Notice that this choice of initialization relies also on the utilization of odd activation functions in deep layers.
\paragraph*{Ground state search.} Following the random initialization we perform a ground state search with Stochastic Reconfiguration with a Hamiltonian tailored to the desired initial state. The termination condition for this ground state search is that the energy variance density of the state is less than $10^{-7}$.

\subsection{Adaptive time step}
We use a second order consistent numerical integration scheme and estimate errors based on varying step sizes. This reduces the required number of Monte Carlo samplings compared to higher order adaptive schemes like the Dormand-Prince method.

As second order consistent integrator we choose the ``Heun method'': Considering an ODE $\dot y=f(y)$ and $y_n=y(t)$, we compute $y_{n+1}=y(t+\tau)$ with
\begin{align}
	k_1&=f(y_n)\ ,\nonumber\\
	k_2&=f(y_n+\tau k_1)\ ,\nonumber\\
	y_{n+1}&=y_n+\frac \tau2(k_1+k_2)\ .
\end{align}
Based on this we can estimate the integration error using varying step sizes $\tau$. If we denote the exact solution by $y(t)$, an integration step with step size $\tau$ yields
\begin{align}
	y_{n+1}=y(t+\tau)+c\tau^3
\end{align}
with an unknown constant $c$, because our integration scheme has an error of $\mathcal O(\tau^3)$.
Alternatively, we can take two steps of size $\tau/2$, resulting in
\begin{align}
	y_{n+1}'=y(t+\tau)+\underbrace{2c\Big(\frac \tau2\Big)^3}_{\delta}
\end{align}
with the integration error $\delta$.
The difference of both solutions is
\begin{align}
	\Delta y_{n+1}=||y_{n+1}-y_{n+1}'||=\Big|\Big|\frac{3}{4}c\tau^3\Big|\Big|=6||\delta||
\end{align}
Given a desired tolerance $\epsilon$ we can adjust the step size based on this to be
\begin{align}
	\tau'=\tau\bigg(\frac{\epsilon}{||\delta||}\bigg)^{1/3}\ .
\end{align}

The choice of a suited norm is a degree of freedom in this procedure. Since in our case the $S$-matrix is the metric tensor of the variational manifold, we employ the norm induced by the $S$-matrix, $||x||_S=\frac{1}{P}\sqrt{\sum_{k,k'} S_{k,k'} x_k^*x_{k'}}$, for that purpose, meaning that we weigh integration errors by their significance for the physical state. When computing the norm, we normalize by the size of the update vectors, i.e., the number of parameters $P$.

\begin{figure}[t]
\includegraphics[width=\columnwidth]{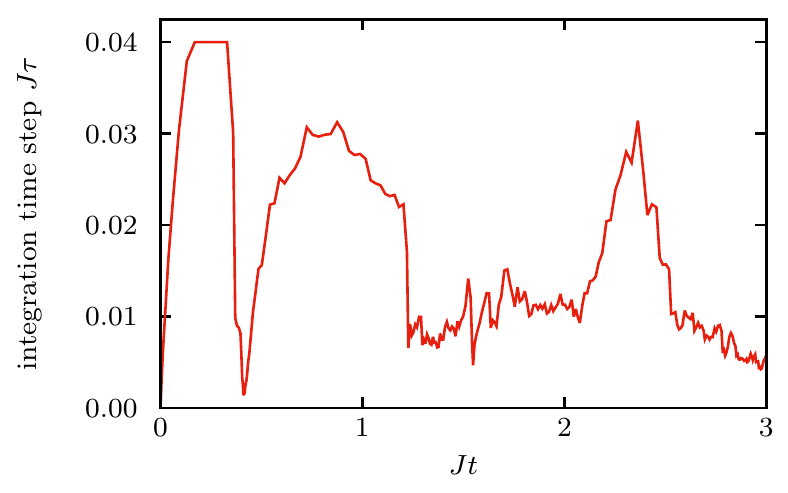}
\caption{Exemplary evolution of the adaptively chosen integration time step $\tau$ during the simulation of the quench to $h=h_c/10$ with $\alpha=(8;10)$.}
\label{fig:timestep}
\end{figure}

An exemplary evolution of the time step during a simulation is shown in Fig.\ \ref{fig:timestep}. Since the step can vary over one order of magnitude, adjusting to the maximal possible value at each time substantially reduces the computational cost.

\subsection{Monte Carlo sampling}
We perform simple Markov Chain Monte Carlo with single spin flip updates to sample $|\psi(s)|^2$. Moreover, every 200th proposed update is a global spin flip to avoid sampling only one of two $\mathbb Z_2$-related modes of the distribution.

\begin{table*}[ht]
\center
\begin{tabular}{|c|c|c|c|c|c|}
\hline
Figure&Network size $\alpha$&
Integration tol. $\epsilon$&
SNR cutoff&
Number of samples $N_{MC}$&
Number of realizations
\\\hline\hline
2a)&$(1;10)$&$5\times10^{-5}$&4&$8\times10^{4}$&1\\
&$(8;10)$&$5\times10^{-5}$&5&$5\times10^{5}$&1\\
&$(4,3,2;4)$&$5\times10^{-5}$&4&$10^{6}$&1\\
&$(4,3,2;6)$&$5\times10^{-5}$&4&$10^{6}$&1\\
\hline
2b)&$(1;10)$&$10^{-4}$&5&$5\times10^{5}$&5\\
&$(8;10)$&$10^{-4}$&8&$5\times10^{5}$&5\\
&$(4,3,2;4)$&$10^{-4}$&8&$2\times10^6$&1\\
&$(4,3,2;6)$&$2.5\times10^{-4}$&8&$5\times10^{5}$&1\\
&$(5,4,3;6)$&$10^{-4}$&8&$5\times10^{5}$&1\\\hline
2c)&$(1;10)$&$5\times10^{-4}$&2&$1.6\times10^{5}$&1\\
&$(8;10)$&$5\times10^{-4}$&4&$3.2\times10^{5}$&1\\
&$(4,3,2;4)$&$5\times10^{-4}$&2&$2\times10^6$&1\\
&$(4,3,2;6)$&$5\times10^{-4}$&2&$10^6$&1\\
&$(5,4,3;6)$&$5\times10^{-4}$&2&$10^6$&1\\\hline
\end{tabular}
\caption{Hyperparameters used for the different simulations presented in the main text.}
\label{tab:hyperparameters}
\end{table*}
\subsection{Regularization scheme}
As it was mentioned in the main text, we regularize the update vectors $\dot{\tilde\eta}_k=\sigma^{-1}_{k}\rho_{k}$ based on the SNR of the Monte Carlo estimate of $\rho_k$. This adds another hyperparameter to the simulation, namely the SNR cutoff $\lambda_{SNR}$. In combination with the adaptive integrator it is beneficial to avoid hard cutoffs and use soft cutoffs instead. Hence, we compute the update vectors as
\begin{align}
	\dot{\tilde\eta}_k=\frac{\sigma^{-1}_{k}\rho_{k}}{1+\Big(\frac{\lambda_{SNR}}{SNR(\rho_k)}\Big)^6}
\end{align}
For this purpose, we estimate the SNR based on the given Monte Carlo sample of size $N_{MC}$ as
\begin{align}
	SNR(\rho_k)=\frac{|\rho_k|\sqrt{N_{MC}}}{\sqrt{\langle |Q_k^*\Elocb|^2\rangle-|\rho_k|^2}}
\end{align}
Clearly, if the Gaussian approximation is accurate, Eq. \eqref{eq:snr} provides a cheaper estimate of the SNR, but the computational cost of the expression above was never relevant in our simulations.
In cases, where the independence of individual samples is doubtful, the fluctuations can alternatively be estimated via a binning analysis. This was, however, not the case in the presented simulations.

\begin{figure}[hb]
\includegraphics[width=\columnwidth]{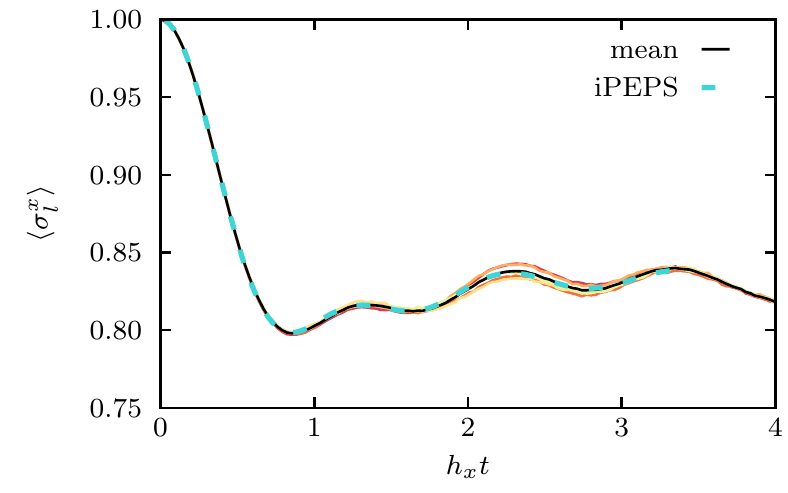}
\caption{Ensemble of five realizations with different random initialization (colored lines) and resulting mean in comparison with the iPEPS reference data for the quench to the critical point, $h=h_c$, and $\alpha=(1;10)$.}
\label{fig:ensemble_avg}
\end{figure}
\subsection{Ensemble averaging of initial conditions}
Despite the high accuracy of the initial ground state search we found for the quench to the critical point that details of the subsequent dynamics can show noticeable fluctuations with varying realizations of the random initialization when the network size is small. For averages over ensembles of initial conditions, however, we found very good agreement with the iPEPS reference data as shown exemplarily in Fig.~\ref{fig:ensemble_avg}

\subsection{Summary of simulation parameters}
In Tab. \ref{tab:hyperparameters} we summarize the various hyperparameters used for the simulations shown in Fig. 2 of the main text. The data in Fig. 1 of the main text was obtained without the SNR-based regularization. Instead Tikhonov regularization was used in combination with the generalized cross-validation criterion to determine an adaptive regularization parameter.

The varying values of the different hyperparameters in Tab. \ref{tab:hyperparameters} might give the impression that precise fine-tuning is required. This is, however, not the case. The variations reflect our experimentation with different settings, which nonetheless led to consistent results.

\begin{figure*}[t]
\includegraphics{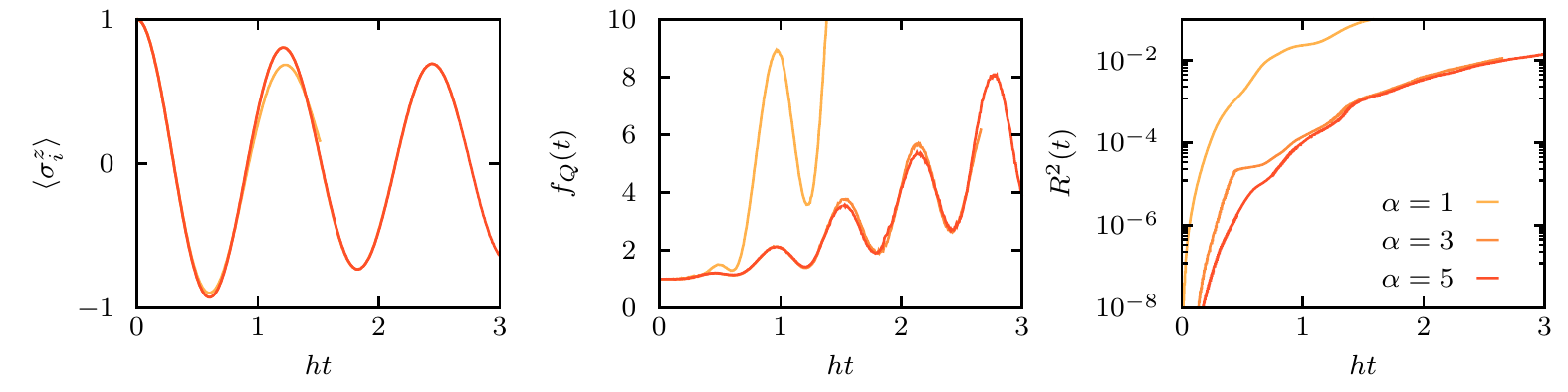}
\caption{Additional data for the quench from the ferromagnetic product state to $h=2.633h_c$ with different network sizes.}
\label{fig:coll_rev_app}
\end{figure*}
\section{Convergence checks}
\subsection{Network size}
A key feature of the neural network approach is the ability to systematically test the accuracy of simulations by comparing results with different network sizes and architectures. Here we provide additional data for the example of Fig. 1 of the main text.
In Fig. \ref{fig:coll_rev_app} we show results for the time evolution of the order parameter $\braket{\sigma_i^z}$, Fisher information density $f_Q(t)$ and the integrated TDVP error $R^2(t)$ obtained with fully connected single layer CNNs of the size $\alpha=1$, $\alpha=3$, and $\alpha=5$. The results for the order parameter fully coincide for $\alpha=3$ and $\alpha=5$, whereas there are small deviations in $f_Q(t)$. Going from $\alpha=1$ to $\alpha=3$ substantially reduces the TDVP error, which is also reflected in both observables. Remarkably, the inaccurate result obtained with the smallest network exhibits a much larger Fisher information density, which means that multi-body entanglement is by far overestimated. This behavior is in contrast to tensor network simulations, which systematically underestimate entanglement when the expressivity is insufficient.

\begin{figure}[b]
\includegraphics{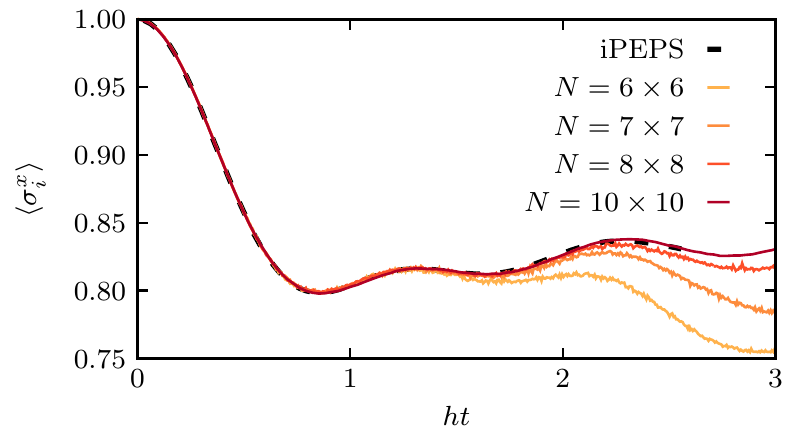}
\caption{Dynamics of transverse magnetization after quenching from the paramagnetic product state to the critical point, $h=h_c$, for different system sizes.}
\label{fig:finite_size_app}
\end{figure}
\subsection{Finite size effect for quench to critical point}
For the quench to the critical point our simulations reveal that the finite system size affects the dynamics of local observables already on timescales that are exceeded by the iPEPS simulation if the system is smaller than $N=8\times8$. In Fig. \ref{fig:finite_size_app} we show results for the dynamics of transverse magnetization after quenching from the paramagnetic product state to the critical point, $h=h_c$, for system sizes $N=6\times6$, $N=7\times7$, $N=8\times8$, and $N=10\times10$. The time interval of agreement with the iPEPS data, which corresponds to dynamics in an infinite system, is extended systematically with increasing system size.

\section{Computational complexity and parallel compute performance}
Variational time evolution with neural network states can benefit substantially from massively parallel compute resources. Here, we outline the complexity of the computationally heavy parts, before sketching our parallelization scheme.

\begin{figure}[ht]
\includegraphics[width=\columnwidth]{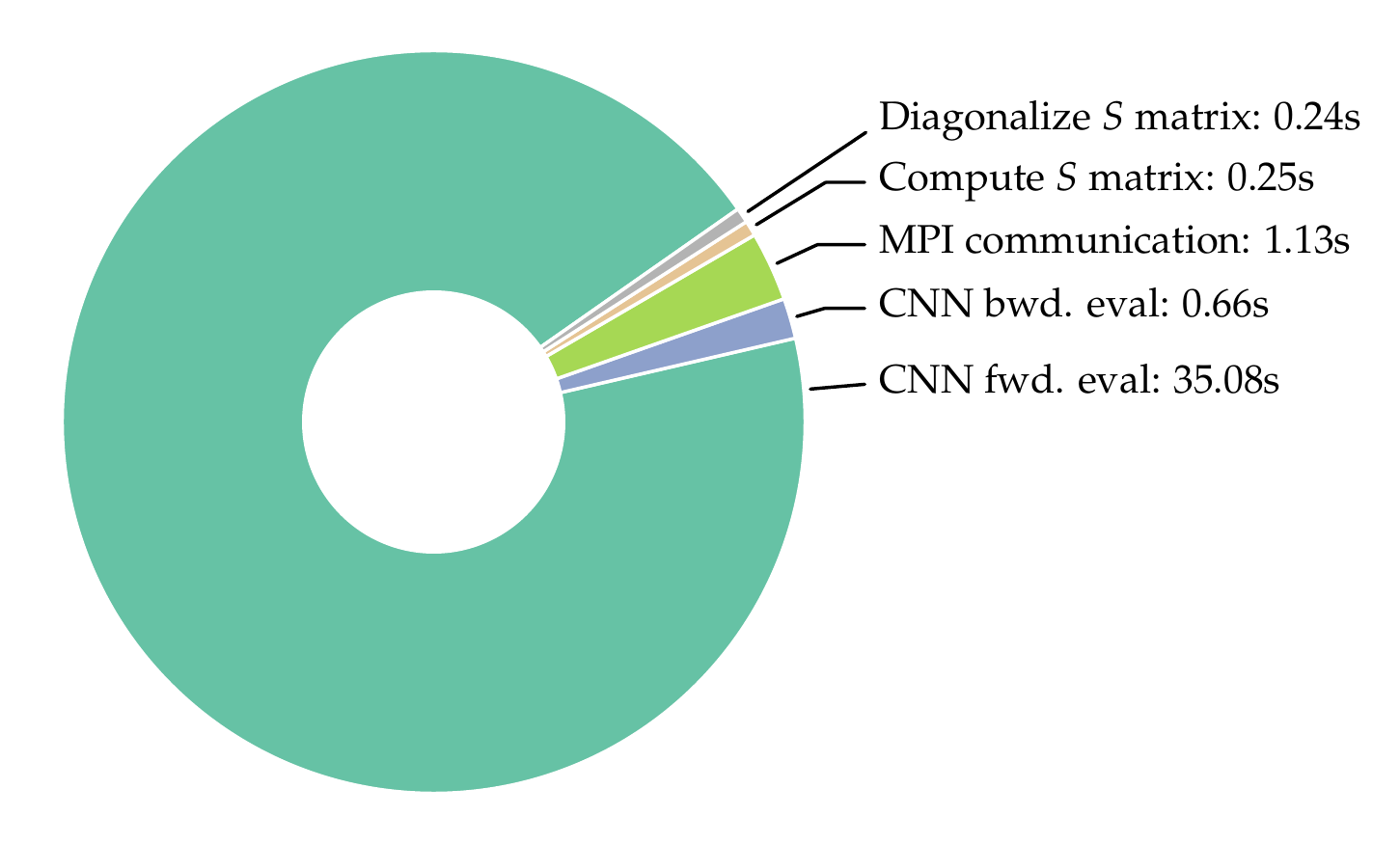}
\caption{Distribution of compute time for one time step with $N_{MC}=5\times10^5$, $P=792$, and $N=100$, using 40 NVIDIA Tesla V100 GPUs. The vast majority of compute time is spent on forward evaluations of the neural network.}
\label{fig:time_pie}
\end{figure}

The algorithm comprises three parts, which can potentially contribute the majority of the computational cost. Which one of these is the computationally most intense part depends on specific number of network parameters $P$, system size $N$ and number of Monte Carlo samples $N_{MC}$. 
\begin{enumerate}
\item \textit{Monte Carlo sampling / evaluating $E_{loc}$:} To this end, we assume that one Monte Carlo sweep consists of $\mathcal O(N)$ proposed updates and that the computation of acceptance ratios comes at the cost of $\mathcal O(1)$ network evaluations. The computational cost to evaluate the CNN on one spin configuration is of $\mathcal O(NP)$ (notice that full network evaluations can in principle be avoided when computing acceptance ratios for few spin-flip updates, see appendix of Ref. [31] of the main text). Hence, the overall cost of Monte Carlo sampling is of $\mathcal O(N_{MC}N^2P)$. The cost of computing the local energy from $N_{MC}$ Monte Carlo samples scales in the same way if the Hamiltonian has $\mathcal O(N)$ off-diagonal contributions, which is the case for local Hamiltonians.
\item \textit{Computing $S$:} To compute the $S$-matrix requires the evaluation of $\mathcal O(P^2)$ correlation functions using $N_{MC}$ samples. Hence the complexity is of $\mathcal O(N_{MC}P^2)$.
\item \textit{Inverting $S$:} The inversion of $S$ boils down to matrix diagonalization. Matrix eigendecomposition of a $P\times P$ matrix has the computational complexity $\mathcal O(P^3)$.
\end{enumerate}
Since in practice $N_{MC}>P$, the overall complexity is $\mathcal O(N_{MC}\times\text{max}(N^2,P)\times P)$. An exemplary distribution of compute time using our GPU parallelized algorithm is shown in Fig.\ \ref{fig:time_pie}, revealing that the majority of the compute time is spent on network evaluations.

\begin{figure*}[hb!]
\includegraphics[width=\textwidth]{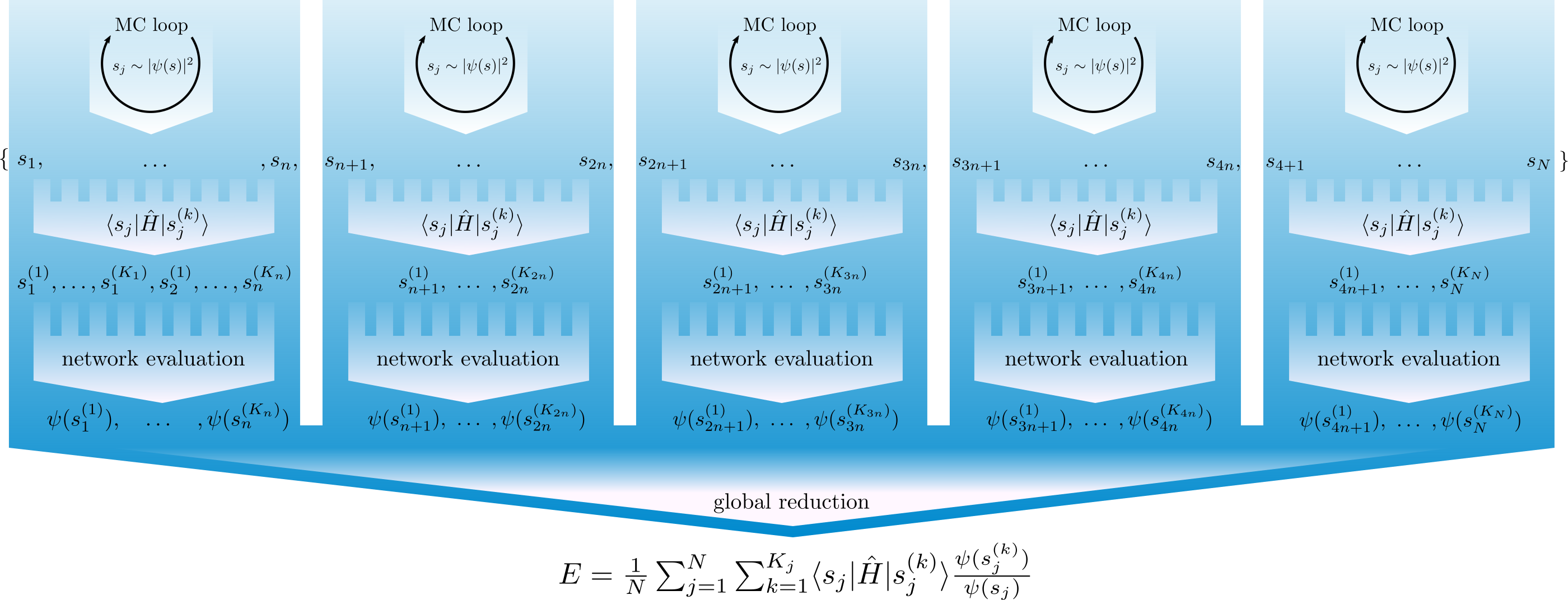}
\caption{Schematic visualization of the parallel implementation of the time evolution algorithm. The top level parallelism of independent Monte Carlo chains can be exploited on a distributed memory machine using a message passing scheme. On the lower level computing matrix elements and evaluating the network for large numbers of input configurations can be parallelized with shared memory using multiple cores of a CPU or GPU accelerators.}
\label{fig:parallel_scheme}
\end{figure*}

Each of the factors appearing in the complexity is amenable to alleviation by parallel execution. In particular, the algorithm allows for a hybrid parallelization that exploits distributed memory parallelism of multiple compute nodes as well as shared memory resources available on individual processors. For the example of computing the energy expectation value the hierarchy of parallelism is schematically depicted in Fig. \ref{fig:parallel_scheme}.
On the top level the Markov Chain Monte Carlo sampling can be distributed over independent processors of a distributed memory machine, exploiting the independence that is inherent to Monte Carlo sampling. For each sample $s_j$ the off-diagonal matrix elements $\braket{s_j|\hat H|s_j^{(k)}}$ and the connected basis configurations $s_j^{(k)}$ have to be determined, which constitutes a set of independent operations that can be carried out in parallel on individual cores of a processor or on a GPU. Finally, the wave function amplitudes have to be determined for all configurations $s_j$ and $s_j^{(k)}$. This corresponds to a large number of independent network evaluations, where, again, the operations of each individual evaluation are well suited for parallelization on a shared memory unit, especially a GPU. Global communication is only required once in order to perform the sum over all matrix elements. A particular measure that enhances the compute performance also for serial execution is network evaluation on batches of input configurations, which increases the arithmetic intensity.

\begin{figure*}[h!]
\includegraphics{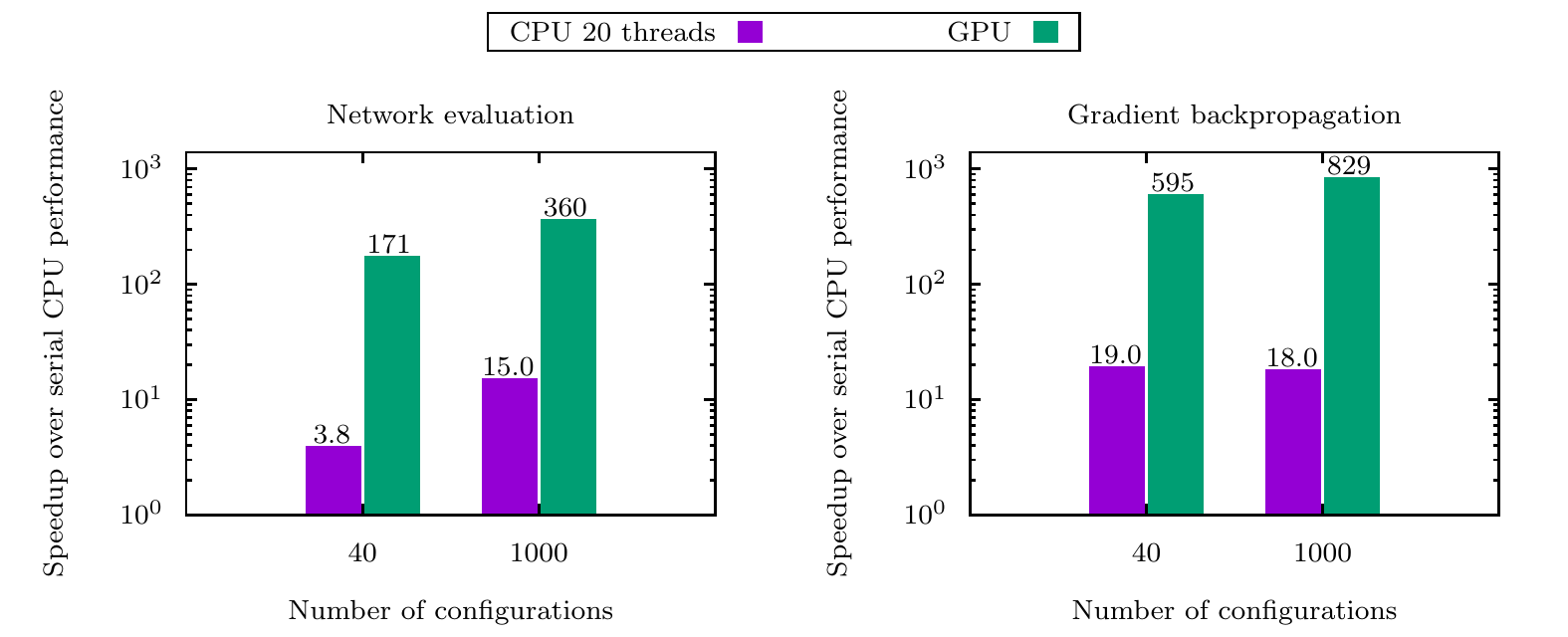}
\caption{Acceleration of the network evaluation and gradient backpropagation using a GPU in comparison with the OpenMP parallelized version on a system of size $N=7\times7$ and with $\alpha=(5,4,3;7)$. Two typical numbers of configurations were considered for batched network evaluation, namely 40 for MC sampling and 1000 for the evaluation of expectation values. The CPU timings were obtained on an \texttt{Intel Xeon E5-2680} and the GPU was an \texttt{NVIDIA V100}.}
\label{fig:parallel_performance}
\end{figure*}

We realized the parallelization on the level of distributed memory using an implementation of Message Passing Interface (MPI) and found perfect speedup with up to 256 processes when taking $N=8\times10^4$ samples. On the shared memory level we used OpenMP/MKL for a CPU implementation and CUDA to alternatively utilize GPU accelerators. In Fig. \ref{fig:parallel_performance} we show the speedup of both implementations obtained over the serial performance for the network evaluation and for gradient backpropagation using the test case of a square lattice with 49 spins and a CNN of size $\alpha=5,4,3$. Our OpenMP/MKL implementation clearly falls behind perfect scaling when using 20 threads. By contrast, both operations are hugely accelerated when using the GPU. In particular, the GPU implementation is 25 times faster than our OpenMP/MKL implementation on 20 cores and still 18 times faster than an ideal parallelization on 20 CPU cores.

